# Studies on Novel Yb-based Candidate Triangular Quantum Antiferromagnets: $Ba_3YbB_3O_9$ and $Ba_3YbB_9O_{18}$

Hwanbeom Cho[1], Stephen J. Blundell[1], Toni Shiroka[2], Kylie MacFarquharson[1], Dharmalingam Prabhakaran[1], and Radu Coldea[1]

[1]*Clarendon Laboratory & Physics Department, University of Oxford, Parks Road, Oxford OX1 3PU, United Kingdom*

[2]*Laboratory for Muon Spin Spectroscopy, Paul Scherrer Institute, 5232 Villigen, Switzerland*

**Abstract**

Yb-based triangular magnets have recently attracted attention as promising candidates to explore frustrated quantum magnetism. However, some candidates have turned out to have significant amounts of site disorder, which significantly affects the low-temperature magnetic behavior. To overcome this issue, Yb-based frustrated systems without structural disorder are required. In this paper, we report physical properties of two Yb-based triangular magnets $Ba_3YbB_3O_9$ and $Ba_3YbB_9O_{18}$, without structural disorder. Via magnetic susceptibility, magnetization, and muon spin rotation measurements, we verified that both $Ba_3YbB_3O_9$ and, particularly studied for the first time, $Ba_3YbB_9O_{18}$ do not show long range magnetic order or glass-like spin freezing down to 0.28 K, but show typical paramagnetic behavior.

**Introduction**

The Yb-based triangular antiferromagnet $YbMgGaO_4$ has attracted great interest recently as a candidate quantum spin liquid [1]. Here the Kramers doublet ground state of $Yb^{3+}$ ions induced by the interplay of strong spin-orbit coupling and octahedral crystal electric field (CEF) leads to magnetic moments with an effective spin-1/2. The strong spin-orbit coupling, furthermore, induces anisotropic exchange interactions, which have been proposed to potentially stabilize a spin-liquid ground state. The small energy scale of exchange interactions of order of 1 K, moreover, has enabled to study the excited states by applying accessible high magnetic field. However, as $YbMgGaO_4$ has site disorder between the $Mg^{2+}$ and $Ga^{3+}$ ions, there is a likely possibility that the absence of long-range magnetic order actually originates from a random singlet pair state stabilized by the randomized exchange interactions [2], with liquid-like signatures in experiments. To overcome the disorder issue, rare-earth frustrated systems without structural disorder are required in order to demonstrate 'pure' quantum spin-liquid behavior [3,4].

$Ba_3YbB_3O_9$ [5] and $Ba_3Yb_9O_{18}$ [6] are novel systems, where the $Yb^{3+}$ ions form hexagonal lattices [see Fig. 1]. The spacing between neighboring Yb ions in their *ab* plane is more than 1 Å shorter than the spacing out of plane, suggesting the systems may be quasi two-dimensional (2D) triangular magnets. In particular, Ba$_3$YbB$_3$O$_9$ has recently been proposed to be a candidate spin liquid based on NMR measurements [7]. In that work, even though the exchange interaction based on Curie-Weiss temperature $\theta_{CW}$ is estimated to be −2.3 K, Ba$_3$YbB$_3$O$_9$ showed neither magnetic ordering nor spin freezing down to the base temperature of 0.26 K. The spin-lattice relaxation rate saturates below 50 K, which denotes that the spins fluctuate fast at base temperature rather than being static or frozen in a spin-glass state.

However, the previous observation can stem from a trivial paramagnetic state, where the thermal fluctuations dominate over the exchange interactions. In rare-earth systems, the energy differences between crystal field levels can be comparable to (or smaller than) the energy scale of room temperature (~30 meV). Therefore, the excitation between the levels can affect the magnetic susceptibility [8], and the results of Curie-Weiss (C-W) fitting vary significantly depending on the temperature region fitted. In this case, it is hard to reliably estimate the average value of exchange interactions from $\theta_{CW}$ without taking into account quantitatively the effect of CEF levels.

Moreover, the energy scale of the interaction implied by the above value of $\theta_{CW}$ seems too large considering the lattice structure. In the case of YbMgGaO$_4$ [9] and NaYbO$_2$ [10], the distance between neighboring Yb$^{3+}$ ions in $ab$ plane is ~3 Å and the energy scale of interactions is of order 1 K. However, the Yb-Yb spacing in Ba$_3$YbB$_3$O$_9$ is significantly larger, ~5 Å, which is too large to have the same magnitude exchange interactions of 1 K. Therefore, the evidence given by the previous NMR study may not be is sufficient to reveal the true nature of the ground state, and further studies at lower temperatures are necessary to identify the ground state. Moreover, to our knowledge, there has been no report yet on the magnetic properties of Ba$_3$YbB$_9$O$_{18}$.

In this paper, we introduce novel hexagonal magnets Ba$_3$YbB$_3$O$_9$ and Ba$_3$Yb$_9$O$_{18}$ as candidates for frustrated quantum magnets, and examine their ground states by studying their properties at temperature down to 0.28 K.

**Methods**

1. Sample synthesis

Polycrystalline Ba$_3$YbB$_3$O$_9$ samples were synthesized via a solid-state reaction method. Initial raw materials of BaCO$_3$, Yb$_2$O$_3$, and B$_2$O$_3$ were mixed in a molar ratio of 6:1:3.15 and then pelletized. Since B$_2$O$_3$ (and H$_3$BO$_3$) absorbs moisture and it stuck to the wall of mortar and pellet die, we need to compensate for the lost B$_2$O$_3$ (H$_3$BO$_3$) by adding more. This is why the ratio of raw materials of Ba$_3$YbB$_3$O$_9$ (and Ba$_3$YbB$_9$O$_{18}$) are slightly deviated from the stoichiometric ratio. The pelletized mixture, contained in a Pt crucible, was sintered at 600 °C in the air for 12 hours. The annealed batch was reground, pelletized, and sintered by increasing temperature from 700 to 900 °C at 100 °C steps with a duration of 12 hours at each step. Finally, the mixture was heated at 1000 °C for 24 h. The grains of polycrystalline samples have thin hexagonal-plaque-like shape, perpendicular to the $c$ axis. This morphology orients additional ~20 % of grains along the [0 0 1] direction. Ba$_3$YbB$_3$O$_9$ does not show congruent melting [11], but it starts to decompose at 1260 °C.

To synthesize polycrystalline $Ba_3YbB_9O_{18}$ samples, we mixed $BaCO_3$, $Yb_2O_3$, and $H_3BO_3$ in a molar ratio of 6:1:18.9. The pelletized mixture, contained in a Pt crucible, was sintered in the air from 600 to 800 °C at 100 °C steps for 12 hours at each step. The annealed batch was reground, pelletized, and sintered by increasing temperature at 900, 960, and 1000 °C steps with a duration of 24 hours at each step. As $Ba_3YbB_9O_{18}$ melts congruently [6], single crystals can be grown by melting the polycrystalline samples in a Pt crucible at 1080 °C, homogenize for 24 h, and then slowly (2 °C/h) cooled down. The transparent $Ba_3YbB_9O_{18}$ single crystals, which are sized up to 5 mm, have morphology of hexagonal plaquettes.

2. X-ray diffraction (XRD)

Crystal structures of the polycrystalline and single-crystal samples were identified by powder and single-crystal x-ray diffraction (see Figs. 2 and 3). Powder diffraction patterns were collected using a Panalytical X'pert PRO powder diffractometer with Cu anode (wavelength $\lambda$ = 1.540598 Å), and an Oxford Diffraction SuperNova diffractometer with Mo source ($\lambda$ = 0.71073 Å) was used to measure a single crystal grain of diameter ~100 μm. The grain size of the powder samples was large enough to show the preferred orientation effect [12], perpendicular to the $c$ axis. The total number of independent reflections given by the single-crystal XRD was 706 (716) for the $Ba_3YbB_3O_9$ ($Ba_3YbB_9O_{18}$) measurements. The observed intensity $F^2_{obs}$, and the calculated intensity, or squared structure factor, $F^2_{calc}$ of each reflection, are well-matched with each other, and this leads to reasonably small agreement factors: $R_{F2}$, $R_{wF2}$, and $R_F$ (Tables 1 and 2). The whole analysis was done through the Rietveld and the integrated intensity refinement methods implemented in the *FullProf* package [12]. The good agreement between the experimental and calculated diffraction patterns is illustrated for the ($h\ k\ 0$) and ($0\ k\ l$) planes in Fig. 3. Furthermore, the orientation of $Ba_3YbB_9O_{18}$ single crystals was also identified using x-ray Laue diffraction.

3. Magnetic measurements

We measured the temperature dependance of the DC magnetic susceptibility $\chi(T)$ by applying a persistent magnetic field of 2000 Oe, and the field dependence of the magnetization $M(H)$ at the base temperature of 2 K, using a Quantum Design PPMS. The magnetization curves of $Ba_3YbB_9O_{18}$ single crystal were measured in two different field directions: perpendicular to the $c$-axis ($H \perp c$) and parallel to the $c$-axis ($H \parallel c$). The mass of each crystal was ~10 mg.

4. Muon spin rotation ($\mu$SR) measurements

$\mu$SR experiments were conducted at the PSI muon source (DOLLY) equipped with a $^3$He cryostat (base temperature of 0.28 K). We first measured the spectra in zero field (ZF) as a function of temperature, with high statistics data sets counted for around 1.5 hours each. At base temperature we collected spectra in several longitudinal fields (LF) up to 4500 G, again for around 1.5 hours each. The $\mu$SR data were analyzed using the $\mu$SR data analysis package, WIMDA [13]. The mass of each batch was ~3 g.

**Results and discussion**

1. Crystal structure

Ba$_3$YbB$_3$O$_9$

Ba$_3$YbB$_3$O$_9$ has a hexagonal crystal structure with space group $P6_3cm$ (#185) and lattice parameters $a$ = 9.3984(2) Å and $c$ = 17.4616(3) [see Fig. 1(a)]. We couldn't detect any evidence of inter-site mixing from analysis of the XRD diffraction data. This structural order can be explained by the large difference in ionic radii of Ba$^{2+}$ (149 pm) and Yb$^{3+}$ (100.8 pm). The hexagonal Yb arrangement is composed of two distinct Yb sites: Yb1 (0 0 0) and Yb2 (2/3 1/3 −0.0050(2)). However, since the fractional $z$ coordinate of the Yb2 is close to zero, the Yb lattice can be regarded as an almost flat 2D plane. Each Yb$^{3+}$ ion is located at the center of oxygen octahedra, with different point group symmetries at the two sites, $C_{3v}$ (Yb1) & $C_3$ (Yb2). The intraplane links between neighboring Yb$^{3+}$ ions are formed via Yb-O-O-Yb paths. The nonmagnetic Ba$^{2+}$ layers isolating the Yb$^{3+}$ layer can lead a strong two dimensionality for the magnetic interactions. The structure is only slightly deviated from a structure with inversion symmetry at the mid-point between in-plane nearest-neighbor Yb-Yb bonds, which can make the Dzyaloshinskii-Moriya (DM) interaction negligible and so simplify the model Hamiltonian of the system.

The crystal structure of Ba$_3$YbB$_3$O$_9$ is somewhat comparable to other recently reported candidate for Yb-based frustrated antiferromagnets, NaBaYb(BO$_3$)$_2$ [14], where YbO$_6$ octahedra form a 2D triangular lattice sandwiched by bilayers made of Na$^+$ and Ba$^{2+}$ ions. Its in-plane exchange paths also follows an Yb-O-O-Yb bond, and the distance between neighboring Yb$^{3+}$ ions (5.3 Å) is comparable to that in Ba$_3$YbB$_3$O$_9$ [5.426(1) Å]. This structural similarity implies that the interactions on the triangular lattice might be similar. However, the out-of-plane distance of NaBaYb(BO$_3$)$_2$ (6.7 Å) is much shorter than that of Ba$_3$YbB$_3$O$_9$ [8.731(1)] Å. This structural difference will make the out-of-plane interactions weaker in Ba$_3$YbB$_3$O$_9$, which leads Ba$_3$YbB$_3$O$_9$ to be closer to an ideal 2D triangular system.

Other detailed structural information obtained from the refinement, such as the fractional atomic positions and thermal parameters, is summarized in Table 1.

Ba$_3$Yb$_9$O$_{18}$

Ba$_3$YbB$_9$O$_{18}$ has a hexagonal lattice with space group $P6_3/m$ (#176), and lattice parameters $a$ = 7.1822(1) Å and $c$ = 16.8869(2) [see Fig. 1(b)]. We couldn't find an evidence of inter-site mixing in this system either. Yb$^{3+}$ ions are located at the center of oxygen octahedra with point group symmetry $C_{3i}$ (= $S_6$). The magnetic Yb$^{3+}$ ions form quasi-two-dimensional triangular layers separated by Ba layers. The longer in-plane Yb-Yb distance of 7.182(1) Å and exchange paths following Yb-O-O-O-Yb bonds imply much smaller interactions. The nearest-neighbor in-plane Yb-Yb bonds have inversion symmetry, which disallows DM interactions on those bonds.

Other detailed structural information obtained from the refinement is summarized in Table 2.

2. Magnetic measurements

Ba$_3$YbB$_3$O$_9$

Inverse of zero-field-cooled (ZFC) and field-cooled (FC) $\chi(T)$ of Ba$_3$Yb$_3$O$_9$ show identical features down to 2 K [Fig. 4(a)]. Absence of an anomaly and bifurcation between ZFC and FC curves indicate that this system does not show long-range order or glass-like spin freezing down to the base temperature studies. The Curie-Weiss (C-W) fitting in the high temperature region gives an effective moment $\mu_{\text{eff}}$ = 4.89(1)$\mu_B$/Yb$^{3+}$, comparable to that of free Yb$^{3+}$ ion (4.54$\mu_B$), and a Curie-Weiss temperature $\theta_{CW}$ = −105(1) K. The deviation of $1/\chi(T)$ from linear behavior below $T$ = 100 K originates most likely from the effect of low-lying field levels and not from magnetic interactions, as the temperature scale is too large for the expected exchange energy scale for rare earth systems. To estimate the energy scale of interactions, we fitted a C-W law to the data, from which the Van Vleck term $\chi_{VV}$ evaluated from $M(H)$ was subtracted, in a temperature region that is low enough to ignore the effect of crystal field excitations (inset of Fig. 3(a)). This gives $\mu_{\text{eff}}$ = 2.56(1)$\mu_B$/Yb$^{3+}$ and $\theta_{CW}$ = −0.24(1) K. Since the ground state of Yb$^{3+}$ is a Kramers doublet $J$ = 1/2, the effective Landé $g$ factor is evaluated to be $g_{CW}$ = 2.96(1). The averaged interactions given by low temperature C-W fit is one order of magnitude smaller than the value previously reported based on the analysis of higher-temperature susceptibility data [7].

The $M(H)$ curve measured at $T$ = 2 K shows a gradual increase, and then saturates to ~1.4$\mu_B$/Yb$^{3+}$ [Fig. 4(b)]. To identify further evidence of magnetic interactions $M(H)$ was fitted to a paramagnetic model (PM) of non-interacting spins with Van Vleck term: $B_J(H) + \chi_{VV} \cdot H$, where $B_J(H)$ is the Brillouin function and $\chi_{VV}$ is a constant Van Vleck susceptibility [Fig. 4(b)]. The overall feature of $M(H)$ can be reproduced [solid curve in Fig. 4(b)] by the paramagnetic model with $g_{PM}$ = 2.77(1) and $\chi_{VV}$ = 3.6(3)·10$^{-7}\mu_B$·Oe$^{-1}$/Yb$^{3+}$, which indicates that the magnetic interaction is too small to be extracted from the $M(H)$ data at 2 K, at this temperature the behavior is similar to that of paramagnetic, non-interacting moments. The $g$ factors: $g_{CW}$ and $g_{PM}$ obtained from different methods are comparable to $g_{avg} = \sqrt{[2(g_\perp)^2 + (g_\parallel)^2]/3}$ = 2.79, reported previously [7].

Ba$_3$YbB$_9$O$_{18}$

The inverse of ZFC and FC $\chi(T)$ of Ba$_3$YbB$_9$O$_{18}$ down to 2 K also shows similar features as observed in Ba$_3$YbB$_3$O$_9$ [Fig. 4(c)]. There is no anomaly or bifurcation, related to the onset of long-range order or spin-glass freezing. The C-W fitting in a high temperature region gives $\mu_{\text{eff}}$ = 4.74(1)$\mu_B$/Yb$^{3+}$ and $\theta_{CW}$ = −99(1) K. The value of $\mu_{\text{eff}}$ is comparable to that of free Yb$^{3+}$ ion supports that the magnetic ion of Ba$_3$YbB$_9$O$_{18}$ is indeed Yb$^{3+}$. The C-W fit on $1/(\chi - \chi_{VV})$ in the low temperature region [inset in Fig. 3(c)] renders the estimation of exchange interactions, $\theta_{CW}$ = −0.077(2) K with $\mu_{\text{eff}}$ = 2.31(1)$\mu_B$/Yb$^{3+}$ and $g_{CW}$ = 2.67(1). Comparing with Ba$_3$YbB$_3$O$_9$, such small interactions of Ba$_3$YbB$_9$O$_{18}$ originate from the longer Yb-Yb distance.

The $M(H)$ of a polycrystalline sample, measured at $T = 2$ K, increases gradually up to a saturated moment of ~1.4$\mu_B$/Yb$^{3+}$ [Fig. 3(d)]. It can be fitted to a paramagnetic model with $g_{PM} = 2.59(1)$ and $\chi_{VV} = 1.1(1) \cdot 10^{-6} \mu_B \cdot $Oe$^{-1}$/Yb$^{3+}$, which indicates that the magnetic interaction is small, and the system is also in the regime of thermal paramagnetism at 2 K. The magnetic moment of a single crystal in an external magnetic field perpendicular to the $c$-axis is larger [$g_\perp = 2.82(1)$] than that in a field along the $c$-axis [$g_\parallel = 2.64(1)$], which indicates that the system has an easy-plane $g$-tensor anisotropy, similar to NaYbO$_2$ [3] and NaYbS$_2$ [4]. This is in contrast to other Yb-based frustrated triangular magnets that have an easy-axis $g$-tensor anisotropy such as YbMgGaO$_4$ [9] and Ba$_3$YbB$_3$O$_9$ [7]. The averaged $g$ factor $g_{avg} = 2.76$ is slightly larger than $g_{CW}$ and $g_{PM}$, which also indicates that the polycrystalline sample has a preferred orientation along the $c$ axis as observed by analyzing powder XRD data.

Comparison of the measured $M(H)$ curve and the expected curve for purely paramagnetic moments gives direct information about the strength of the magnetic interactions. Figure 4(e) compares $M(H)$ of several other Yb-based frustrated antiferromagnets: BaNaYb(BO$_3$)$_2$ [14], YbMgGaO$_4$ [9], and NaYbO$_2$ [10], with corresponding curves for paramagnetic behavior at the same temperature for the same magnitude moments (same $g$ values). The large contrast in the case of YbMgGaO$_4$ and NaYbO$_2$ indicates that exchange interactions are relatively large, but the small deviation in Ba$_3$YbB$_3$O$_9$, BaYbB$_9$O$_{18}$, and BaNaYb(BO$_3$)$_2$ indicates that interactions here are rather small. Indeed, the $\theta_{CW}$ of former ones: YbMgGaO$_4$ (−2.7 K) and NaYbO$_2$ (−5.6 K) are one order of magnitude larger than the latter ones: Ba$_3$YbB$_3$O$_9$ (−0.24 K), BaYbB$_9$O$_{18}$ (−0.077 K), and BaNaYb(BO$_3$)$_2$ (−0.15 K). This tendency relates to the spacing between Yb$^{3+}$ ions in the triangular lattice. In the case of YbMgGaO$_4$ and NaYbO$_2$, it is about 3 Å, but in the other case, it is larger than 5 Å. Therefore, the energy scale of interactions in Ba$_3$YbB$_3$O$_9$ is unlikely to be as large as 1 K, but more likely of the order of 0.1 K as we evaluated from the low-temperature C-W fit.

3. Muon spin rotation ($\mu$SR) experiment

Ba$_3$YbB$_3$O$_9$

A $\mu$SR experiment was conducted to search for any sign of local order and characterize the fluctuations of local moments. Down to $T = 0.28$ K, we could not observe any oscillating features, which would be expected in the presence of long-range magnetic order. The ZF $\mu$SR depolarization spectra [Fig. 5(a)] were fitted to the function

$$A(t) = A_1 \cdot e^{-\lambda_1 t} + A_2 \cdot e^{-\lambda_2 t} + A_{bg}, \quad (1)$$

where $A_1$ and $A_2$ are amplitudes of the two exponentials with relaxation rates $\lambda_1$ and $\lambda_2$, and $A_{bg}$ represents a temperature-independent constant background (which was around 1 %). The amplitudes $A_1$ and $A_2$ were held fixed across the whole temperature range studied (in the ratio approximately 0.65:0.35) and we interpret the two exponentials as arising from two different muon sites in the sample. The temperature dependence of the two relaxation rates is shown in Fig. 5(b), illustrating that both relaxation rates rise on cooling but then saturate at low temperatures, though to different values. This behavior is similar to that seen in the previous NMR study [7]. The exponential decay emerges when the electron moments of Yb$^{3+}$ ions are in the fast fluctuation limit and the relaxation rate in zero field is then given by $\lambda = 2\Delta_e^2/\nu$, where $\Delta_e$ is the width of a distribution of local fields at the muon site and $\nu$ is the fluctuation frequency of the local field at the muon site [15]. Since $\Delta_e$ is not expected to be temperature dependent, the temperature dependence must arise from $\nu$ and we interpret the observed behavior in terms of two contributions: a temperature-independent term (persistent fluctuations) and a thermal term for which fluctuations slow down on cooling. This can be modelled by writing $\lambda^{-1} = \lambda_0^{-1} + C \cdot e^{-\Delta_{CEF}/T}$, where the thermal term is given an activated dependence (an Orbach process [16]) with an energy splitting $\Delta_{CEF}$ that is expected to be the energy of the first-excited crystal-field energy level. The fits yield an energy splitting $\Delta_{CEF}$ = 35(4) meV for $\lambda_1$ and 45(2) meV for $\lambda_2$, but these are consistent with an average value of 40(5) meV, comparable to that evaluated from a spectroscopic method [11].

The application of a longitudinal field has two distinct effects: (1) the sample magnetizes, resulting in both an additional longitudinal field and transverse field at the muon site (due to the anisotropic nature of the dipolar interaction) and (2) the muon becomes decoupled from the fluctuating field due to the $Yb^{3+}$ moments. In addition, the fluctuations themselves may be field dependent. This situation is therefore quite complex and we focus only on the decoupling effect, studying the low-field behavior that was performed at 0.28 K, fitting the data with the amplitudes fixed as in Eq. (1). In this case, the relaxation rate $\lambda = 2\Delta_e^2\nu/(\nu^2 + \gamma_\mu^2 B^2)$ [15,17], where $\gamma_\mu$ is the gyromagnetic ratio of muon (85.1 MHz/kG), and $B$ is the applied longitudinal magnetic field. Plotting $1/\lambda$ against $B^2$ (see Fig. 6) then allows a simple extraction of the key parameters, although the effect of the field-dependent magnetizing of the sample implies that this can only be a rough estimate. These results yield $\nu$ ~80 MHz for the persistent spin fluctuations and $\Delta_{e1}$ ~5 MHz and $\Delta_{e2}$ ~20 MHz, implying that the second exponential corresponds to a muon site that is closer to a $Yb^{3+}$ moment, and also confirming that the fluctuations are in the fast-fluctuation limit ($\nu \gg 5\Delta_e$). The fluctuation rate $\nu$ is comparable to, but somewhat smaller than, that of high temperature mean-field (MF) exchange fluctuations [18,19]: $\nu_{MF} = \sqrt{z} \cdot S \cdot J_0/h = \sqrt{6} \cdot (1/2) \cdot (k_B \theta_{CW}/3)/h$ = 2.0 GHz and this contradicts to $\nu$ of correlated states where $\nu$ becomes three orders of magnitude smaller than $\nu_{MF}$ [18].

In highly frustrated systems, upon decreasing temperature the relaxation rate $\lambda$ is expected to increase down to temperatures comparable to $|\theta_{CW}|$, where thermal fluctuations cease and short-range order starts to emerge, and then it might stabilize around a finite value if spins continue to dynamically fluctuate [10,18,20]. However, $Ba_3YbB_3O_9$ doesn't show such behavior down to 0.28 K, but our results demonstrate that thermal fluctuations still dominate the relaxation rate behavior at that temperature.

$Ba_3YbB_9O_{18}$

Down to $T$ = 0.28 K, the $\mu$SR depolarization spectra of $Ba_3YbB_9O_{18}$ do not show any oscillating feature and so once again we can rule out the presence of long-range magnetic order. The analysis proceeded as for $Ba_3YbB_3O_9$ and the corresponding ZF spectra are shown in Fig. 5(c), also fitted to Eq. (1) with amplitudes fixed and in the ratio $A_1$:$A_2$ ~ 0.55:0.45. Our fits similarly give an Orbach process with energy splitting $\Delta_{CEF} \sim 50(5)$ meV, slightly larger than that of $Ba_3YbB_3O_9$. The dataset for the LF spectra is more limited (see Fig. 6) and do not allow for a precise determination of the fluctuation parameters, but are consistent with a very similar fluctuation rate as observed for $Ba_3YbB_3O_9$. Nevertheless, it is clear that all the way down to the base temperature of our experiment (0.28 K), and in common with $Ba_3YbB_3O_9$, the relaxation rate $\lambda$ of $Ba_3YbB_9O_{18}$ remains dominated by fast, persistent spin dynamics.

**Conclusion**

$Ba_3YbB_3O_9$ and $Ba_3YbB_9O_{18}$ have two-dimensional triangular lattices made of well-ordered $Yb^{3+}$ ions with no structural atomic site mixing. In the octahedral crystal electric field, the Kramers doublet of $Yb^{3+}$ ions gives an effective quantum spin $J = 1/2$ to each system. On top of these characteristics, the negative Curie-Weiss temperature: −0.24 K ($Ba_3YbB_3O_9$) and −0.077 K ($Ba_3YbB_9O_{18}$) of each system satisfies the conditions to be a candidate for frustrated magnets. Via magnetic susceptibility, magnetization, and muon spin rotation measurements, we verified that $Ba_3YbB_3O_9$ and $Ba_3YbB_9O_{18}$ do not show long range order or glass-like spin freezing down to $T = 0.28$ K, but they show typical paramagnetic behavior at the high temperature larger than the exchange interaction. In order to identify any ordered ground state of either $Ba_3YbB_3O_9$ and $Ba_3YbB_9O_{18}$, further experiments at significantly lower temperature are required.

**Acknowledgement**

This work was supported by the European Research Council (ERC) under the European Union's Horizon's 2020 research and innovation program Grant Agreement Number 788814 (H.C., D.P., R.C.) and by EPSRC (UK) under Grant EP/N023803/1 (S.J.B). Part of this work was carried out at the Swiss Muon Source, Paul Scherrer Institut, Switzerland and we are grateful for the provision of beamtime.

**Figures and Tables**

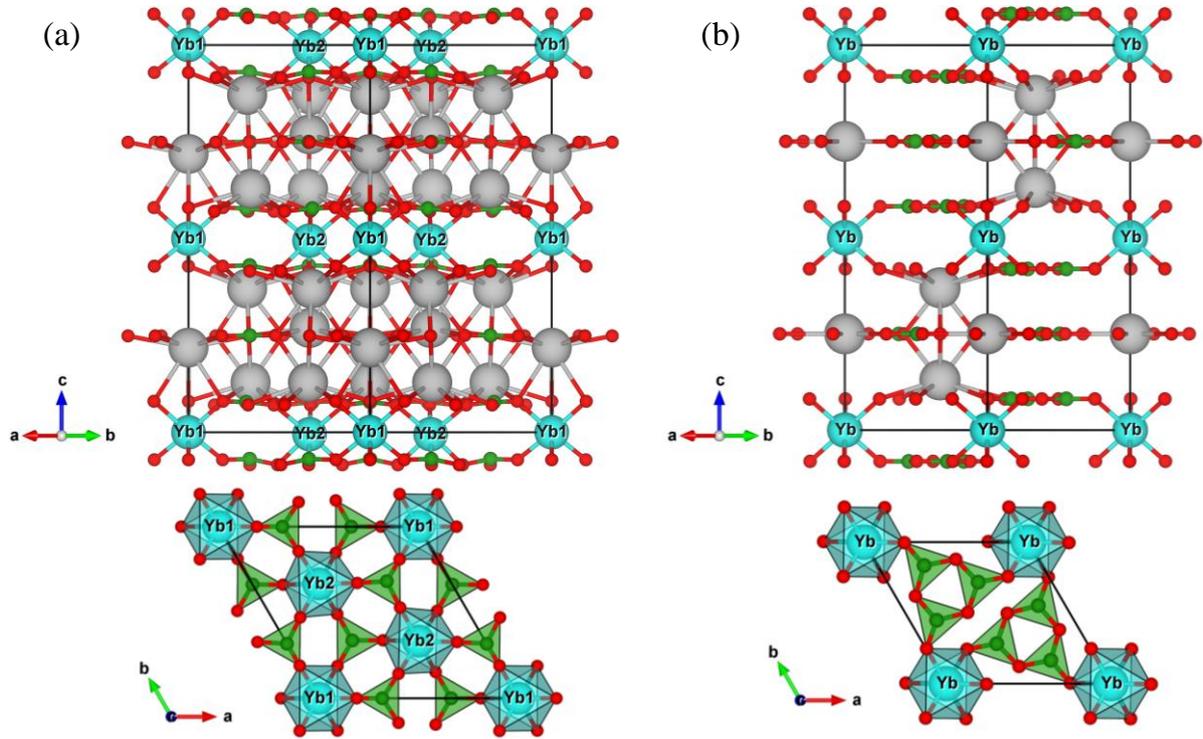

FIG. 1. (a) Crystal structure of $Ba_3YbB_3O_9$ drawn using the crystallographic information given in Table 1. Cyan, grey, red, and green balls denote $Yb^{3+}$, $Ba^{2+}$, $O^{2-}$ and $B^{3+}$ ions, respectively. The Yb layers in the *ab* plane are isolated by Ba layers along *c* axis. The Yb1 and Yb2 sites form a triangular lattice. (b) Crystal structure of $Ba_3YbB_9O_{18}$ from the data in Table 2. Yb sites form triangular lattices in the *ab* plane, separated by Ba layers.

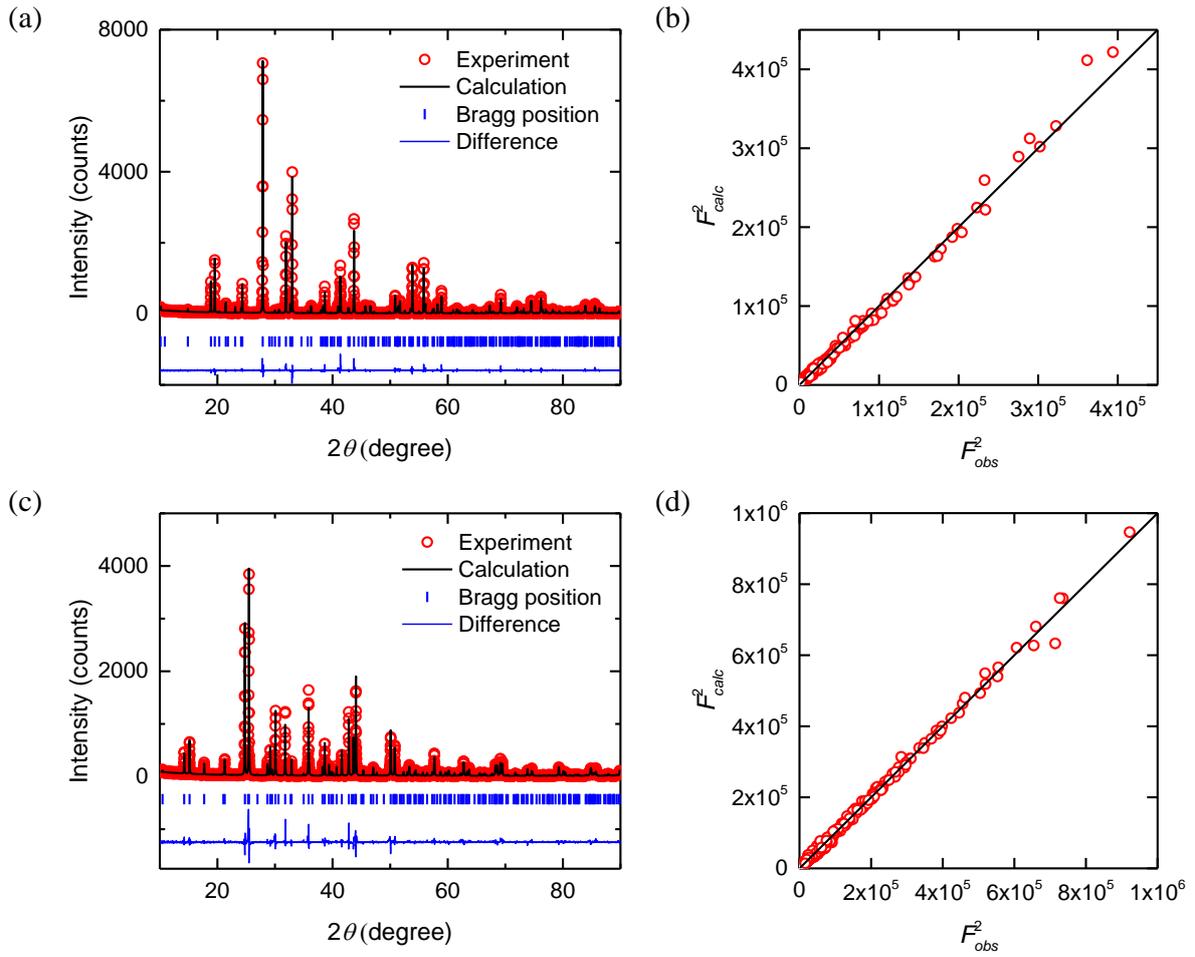

FIG. 2. (a) Powder XRD pattern of $Ba_3YbB_3O_9$. The crystal structure was refined using a model with a preferred orientation ($G_1$ = 0.5623 and $G_2$ = 0.7952 of modified March's function [12]) around [0 0 1]. This preferred orientation originates from the thin hexagonal-plate-like morphology of grains. (b) Refinement result of XRD data from a single crystal grain of $Ba_3YbB_3O_9$ of size ~100 μm. (c) Powder XRD pattern of $Ba_3YbB_9O_{18}$, also refined with a preferred orientation ($G_1$ = 0.3591 and $G_2$ = 0.8576) around [0 0 1]. (d) Refinement result of XRD data from a single crystal grain of $Ba_3YbB_9O_{18}$ of size ~100 μm.

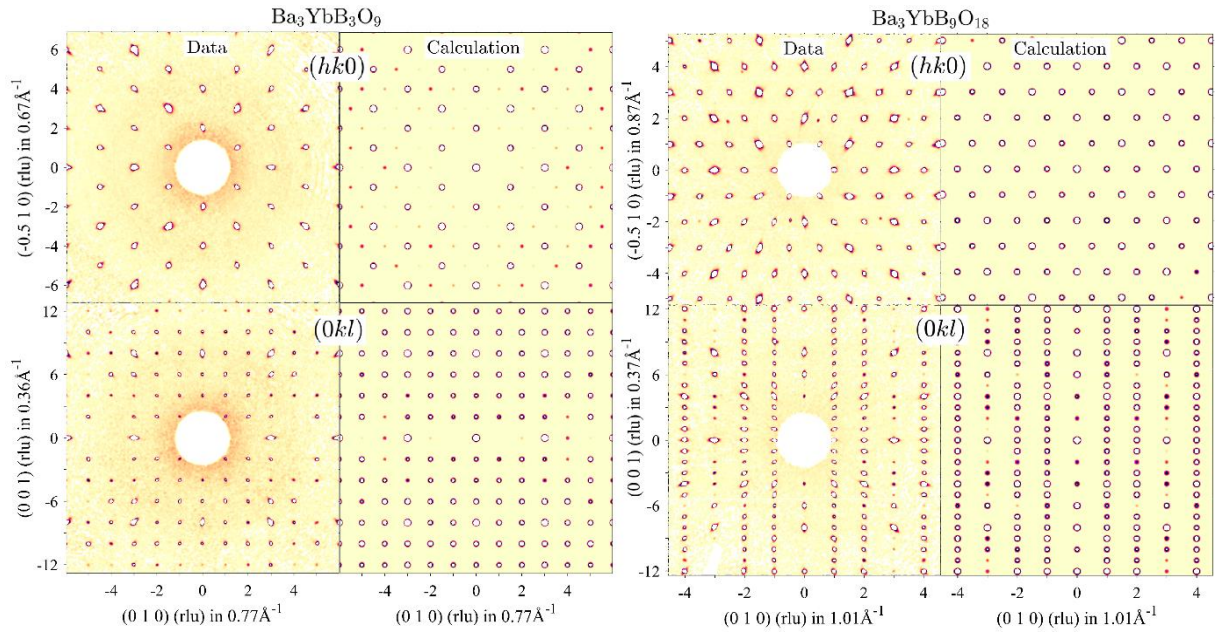

FIG 3. Side by side experimental vs calculated x-ray diffraction patterns in two representative reciprocal lattice planes, ($h\,k\,0$) (top row) and ($0\,k\,l$) (bottom row), for $Ba_3YbB_3O_9$ (left) and $Ba_3YbB_9O_{18}$ (right).

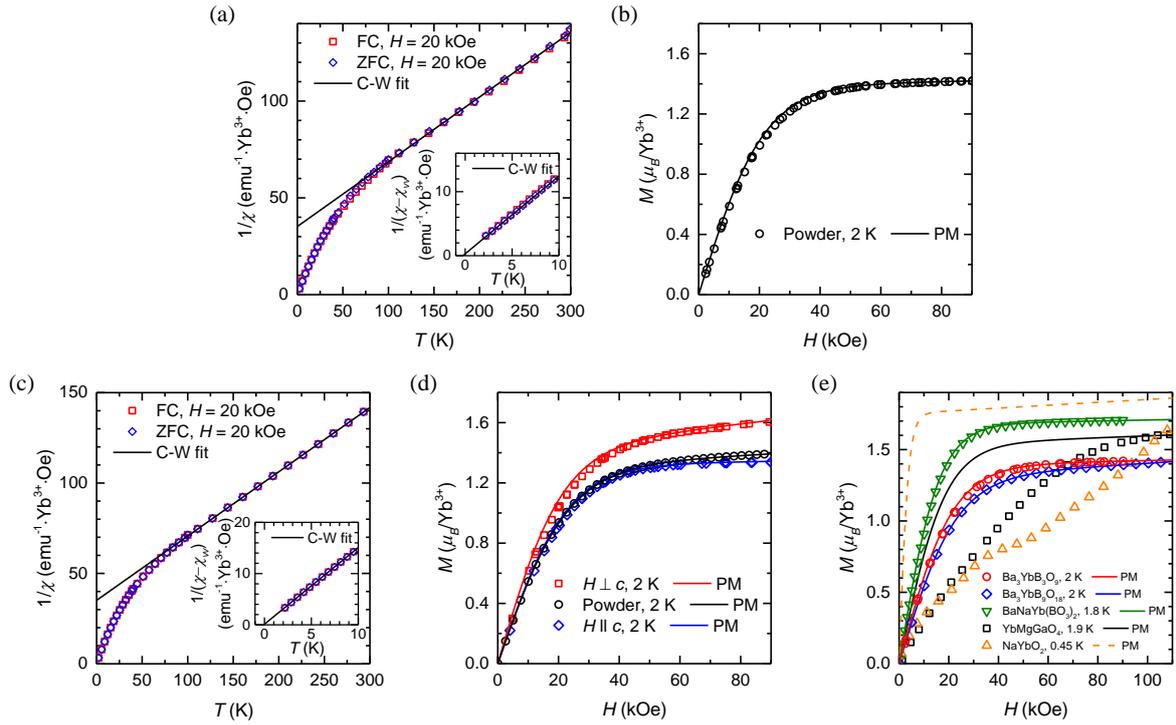

FIG. 4. (a) Inverse of field-cooled (square) and zero-field-cooled (diamond) $\chi(T)$ data of $Ba_3YbB_3O_9$ under external field of $H = 20$ kOe. The data was fitted with Curie-Weiss fit (solid line) in the high temperature region (150-295 K), and the low temperature region (3-7 K, in the inset). (b) $M(H)$ data of $Ba_3YbB_3O_9$ at 2 K, comparable with typical paramagnetic (PM) behavior of non-interacting moments (solid line). (c) $1/\chi(T)$ data of $Ba_3YbB_9O_{18}$ and Curie-Weiss fit (solid line) in the high temperature region (150-295 K), and low temperature region (3-7 K, in the inset). (d) $M(H)$ data of polycrystalline $Ba_3YbB_9O_{18}$ sample at 2 K (circle), and that of crystals under field perpendicular to $c$ axis (square) and parallel to $c$ axis (diamond). Each of them is comparable with the behavior for the non-interacting paramagnetic model (solid lines). (e) $M(H)$ data (symbols) of several Yb-based triangular antiferromagnets: $Ba_3YbB_3O_9$ and $Ba_3YbB_9O_{18}$ (this work), $BaNaYb(BO_3)_2$ [14], $YbMgGaO_4$ [9], and $NaYbO_2$ [10]. Solid curves of the same color as the data points show the expected $M(H)$ curves assuming non-interacting paramagnetic moments.

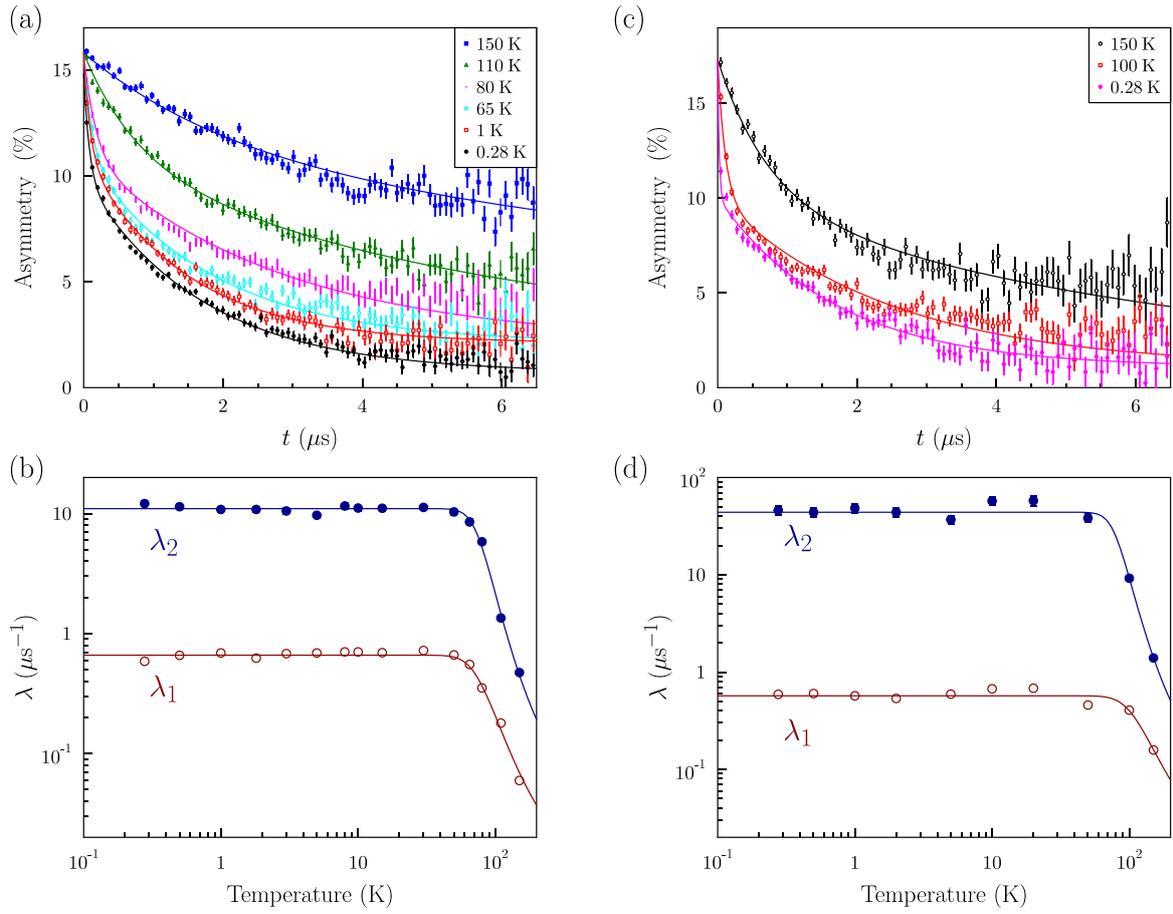

FIG. 5. (a) ZF $\mu$SR depolarization spectra for $Ba_3YbB_3O_9$ at various temperatures. Each spectrum can be fitted to Eq. (1) (solid lines). (b) Temperature dependence of the two relaxation rates $\lambda_1$ and $\lambda_2$, fitted according to a model involving an Orbach process (solid line). (c) ZF $\mu$SR depolarization spectra $Ba_3YbB_9O_{18}$ and (d) the corresponding temperature dependence of the two relaxation rates.

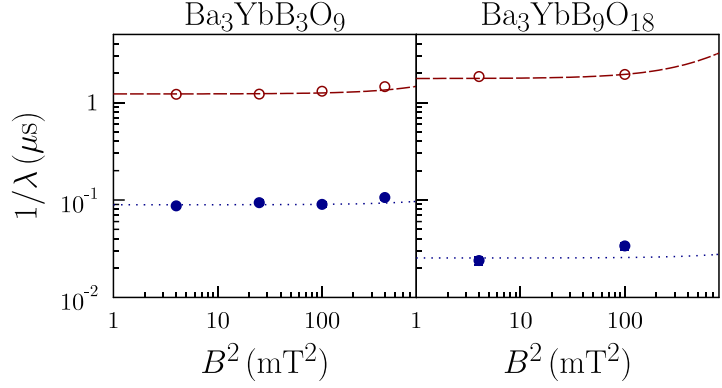

FIG. 6. The data obtained from fits of the low-field LF spectra at 0.28 K for the two compounds yields an estimate of the fluctuation frequency (see text).

Table 1. Structural information of $Ba_3YbB_3O_9$ obtained from refinement methods.

| Atom | Wyckoff symbol | Site symmetry | x/a | y/b | z/c | $U_{iso}$ (Å$^2$) |
|---|---|---|---|---|---|---|
| colspan=7 | Space group: $P6_3cm$ (No. 185), Z=6 |||||||
| colspan=7 | Cell dimensions: a(Å)=9.3984(2), c(Å)=17.4616(3), α=β=90°, γ=120°, V(Å$^3$)=1335.75(4) |||||||
| Yb1 | 2a | 3 m | 0 | 0 | 0.0000 | 0.0097(9) |
| Yb2 | 4b | 3 | 2/3 | 1/3 | −0.0050(2) | 0.0054(5) |
| Ba1 | 2a | 3 m | 0 | 0 | 0.2187(5) | 0.0097(9) |
| Ba2 | 4b | 3 | 2/3 | 1/3 | 0.2692(3) | 0.0054(5) |
| Ba3 | 6c | m | 0.3419(4) | 0.3419(4) | 0.1290(4) | 0.0020(10) |
| Ba4 | 6c | m | 0.3230(5) | 0.3230(5) | 0.3710(4) | 0.0101(14) |
| B1 | 6c | m | 0.3180(60) | 0.3180(60) | 0.5780(30) | 0.0020(30) |
| B2 | 6c | m | 0.3380(110) | 0.3380(110) | 0.7500(50) | 0.0020(30) |
| B3 | 6c | m | 0.3370(60) | 0.3370(60) | 0.9310(30) | 0.0020(30) |
| O1 | 6c | m | 0.1900(40) | 0.1900(40) | 0.5801(19) | 0.0020(30) |
| O2 | 12d | 1 | 0.3280(40) | 0.4710(40) | 0.5706(16) | 0.0080(30) |
| O3 | 12d | 1 | 0.1830(50) | 0.3330(60) | 0.7489(15) | 0.0080(30) |
| O4 | 6c | m | 0.4790(60) | 0.4790(60) | 0.7504(16) | 0.0020(30) |
| O5 | 12d | 1 | 0.3380(40) | 0.4910(40) | 0.9161(13) | 0.0080(30) |
| O6 | 6c | m | 0.1870(50) | 0.1870(50) | 0.9290(20) | 0.0020(30) |
| Atom | $U_{11}$ (Å$^2$) | $U_{22}$ (Å$^2$) | $U_{33}$ (Å$^2$) | $U_{12}$ (Å$^2$) | $U_{13}$ (Å$^2$) | $U_{23}$ (Å$^2$) |
| Ba3 | 0.0026(11) | 0.0026(11) | 0.0009(9) | −0.0011(13) | 0.0001(4) | 0.0001(4) |
| Ba4 | 0.0088(14) | 0.0088(14) | 0.0128(14) | 0.0017(15) | 0.0013(12) | 0.0013(12) |
| colspan=7 | Agreement factors: $\chi^2$=1.44, $R_{F2}$(%)=8.39, $R_{wF2}$(%)=13.1, $R_F$(%)=8.13 |||||||

Table 2. Structural information of $Ba_3YbB_9O_{18}$ obtained from refinement methods.

| | Space group: $P6_3/m$ (No. 176), $Z=2$ | | | | | |
|---|---|---|---|---|---|---|
| | Cell dimensions: $a$(Å)=7.1822(1), $c$(Å)=16.8869(2), $\alpha=\beta=90°$, $\gamma=120°$, $V$(Å$^3$)=754.39(2) | | | | | |
| Atom | Wyckoff symbol | Site symmetry | $x/a$ | $y/b$ | $z/c$ | $U_{iso}$ (Å$^2$) |
| Yb | 2$b$ | $\bar{3}$ | 0 | 0 | 0 | 0.0071(3) |
| Ba1 | 4$f$ | 3 | 2/3 | 1/3 | 0.1302(1) | 0.0098(3) |
| Ba2 | 2$a$ | $\bar{6}$ | 0 | 0 | 1/4 | 0.0442(8) |
| B1 | 6$h$ | $m$ | 0.5060(17) | −0.1155(17) | 1/4 | 0.0119(20) |
| B2 | 12$i$ | 1 | −0.4478(11) | −0.2826(11) | 0.0813(5) | 0.0113(13) |
| O1 | 6$h$ | $m$ | 0.2906(10) | −0.1656(10) | 1/4 | 0.0129(13) |
| O2 | 12$i$ | 1 | 0.4971(7) | −0.1215(7) | 0.0810(3) | 0.0139(10) |
| O3 | 6$h$ | $m$ | 0.6686(10) | 0.0823(10) | 1/4 | 0.0112(13) |
| O4 | 12$i$ | 1 | −0.2472(7) | −0.2433(7) | 0.0810(3) | 0.0137(9) |
| Atom | $U_{11}$ (Å$^2$) | $U_{22}$ (Å$^2$) | $U_{33}$ (Å$^2$) | $U_{12}$ (Å$^2$) | $U_{13}$ (Å$^2$) | $U_{23}$ (Å$^2$) |
| Yb | 0.0065(3) | 0.0065(3) | 0.0083(3) | 0.0032(3) | 0 | 0 |
| Ba1 | 0.0092(3) | 0.0092(3) | 0.0111(3) | 0.0046(3) | 0 | 0 |
| Ba2 | 0.0063(4) | 0.0063(4) | 0.1201(15) | 0.0032(4) | 0 | 0 |
| Agreement factors: $\chi^2$=1.03, $R_{F2}$(%)=4.10, $R_{wF2}$(%)=7.21, $R_F$(%)=3.80 | | | | | | |